\documentclass[referee]{chjaa}          

\usepackage{graphicx,times}             
\input{epsf.sty}                        
\input{psfig.sty}                       

\newcommand{\msun}{$M_{\odot}$}

\newcommand{\Mbh}{{M_{\rm BH}}}

\def\L60{L_{60{\mu}\rm m}}
\def\Lbol{L_{\rm bol}}

\newcommand{\Mdot}{\dot{M}_{\rm acc}}
\newcommand{\Mstardot}{\dot{M}_\star}
\newcommand{\myear}{M_\odot\, {\rm yr^{-1}}}

\begin{document}

   \title{The Growth of Black Holes and Their Host Spheroids in (Sub)mm-loud QSOs at High Redshift
}

   \volnopage{Vol.0 (200x) No.0, 000--000}      
   \setcounter{page}{1}           

   \author{C.N. Hao
      \inst{1,2,3}\mailto{}
   \and X.Y. Xia
      \inst{1}
   \and S. Mao
      \inst{4}
   \and Z.G. Deng
      \inst{5}
   \and Hong Wu
      \inst{2}
      }

   \institute{Center of Astrophysics, Tianjin Normal University, 300384 Tianjin, China\\ 
   \email{hcn@ast.cam.ac.uk}
        \and
             National Astronomical Observatories, Chinese Academy of Sciences, A20 Datun Road, 100012 Beijing, China\\
	\and
             Current address: Institute of Astronomy, University of Cambridge, Madingley Road, Cambridge CB3 0HA, UK\\
        \and   
             Jodrell Bank Observatory, University of Manchester, Macclesfield, Cheshire SK11 9DL, UK\\
        \and 
             College of Physical Science, Graduate School of the Chinese Academy of Sciences, Beijing 100049, China\\
          }

   \date{Received~~2007 month day; accepted~~2007~~month day}

   \abstract{
We study the growth of black holes and stellar population in spheroids 
at high redshift using several (sub)mm-loud QSO samples. 
Applying the same criteria established in an earlier work, we find 
that, similar to IR QSOs at low redshift, the far-infrared 
emission of these (sub)mm-loud QSOs mainly originates from dust heated 
by starbursts. 
By combining low-z IR QSOs and high-z (sub)mm-loud 
QSOs, we find a trend that the star formation rate ($\Mstardot$) increases with 
the accretion rate ($\Mdot$). 
We compare the values of $\Mstardot/\Mdot$ 
for submm emitting galaxies (SMGs), far-infrared 
ultraluminous/hyperluminous QSOs and typical QSOs, and construct a likely evolution 
scenario for these objects. The (sub)mm-loud QSO transition phase has both 
high $\Mdot$ and $\Mstardot$ and hence is important 
for establishing the correlation between the masses of black holes and spheroids.
   \keywords{galaxies: active --- galaxies: evolution --- galaxies: high redshift --- galaxies: interactions --- quasars: general --- galaxies: starburst }
   }

   \authorrunning{C. N. Hao et al. }            
   \titlerunning{The Growth of Black Holes and Their Host Spheroids } 

   \maketitle

%
%
\section{Introduction}           
\label{sect:intro}
~~~~In the last few years, it has become increasingly clear that the growth 
of supermassive black holes and their host spheroids 
must be closely related as the black hole mass has correlations with 
the galactic velocity dispersion (e.g., Ferrarese \& Merritt 2000; Tremaine 
et al. 2002) and the luminosity/mass of the hot stellar component of the 
host galaxy (e.g., Magorrian et al. 1998; Laor 1998; Kormendy \& Gebhardt 2001).

However, it remains unclear how the correlations arise. Although 
much efforts have been made both theoretically 
(e.g. Silk \& Rees 1998; Haehnelt \& Kauffmann 2000; Adams et al. 2001; Burkert \& Silk 2001; 
Balberg \& Shapiro 2002; Springel et al. 2005; 
Di Matteo et al. 2005) and observationally (e.g. Shields et al. 2003; 
Treu et al. 2004; Heckman et al. 2004; Walter et al. 2004; Borys et al. 2005, Shields et al. 2006), 
the definitive interpretation still remains to be 
established. 

In an earlier work, Hao et al. (2005) studied the QSOs/Seyfert 1s selected 
from local ultraluminous infrared galaxies (IR QSOs) and found a correlation 
between the star formation rate ($\Mstardot$) and the accretion rate to 
central AGNs ($\Mdot$) for IR QSOs; the ratio of $\Mstardot$ to $\Mdot$ 
is about several hundred (Hao et al. 2005, hereafter paper I). 
These IR QSOs not only have massive starbursts occurring in their host galaxies 
their optical spectroscopic and X-ray properties also exhibit 
characteristics of young forming QSOs (Zheng et al. 2002). 
Thus IR QSOs may be an important evolution phase 
from massive starbursts to luminous QSOs and later to elliptical 
galaxies. During this transition phase, both the spheroid components 
and central black holes grow rapidly. 
Therefore it is important to extend our previous studies for 
the local universe to high redshift 
in order to better understand how the correlation between the masses of 
spheroids and their central black holes arises and evolves as a function of 
cosmic time.

Massive submillimeter emitting galaxies 
(SMGs), uncovered by deep SCUBA surveys on blank fields, 
resemble scaled-up versions of the local ultraluminous infrared galaxies 
at high redshift (e.g. Tacconi et al. 2006) -- their star formation rates and 
molecular gas reservoirs are one order of magnitude higher 
(on average) than their counterparts at low redshift 
(e.g. Kim et al. 1998; Downes \& Solomon 1998; 
Chapman et al. 2004; Alexander et al. 2005b). 
In addition, recent investigations on SMGs by ultra-deep 
X-ray observations (the 2 Ms Chandra Deep Field North) 
reveal only modest $\Mdot \la 1 \myear$ for their central AGNs 
(Alexander et al. 2005a,b). These observations indicate that 
there are few luminous QSOs in the current SMG samples and hence they are 
not ideal samples from which we can identify high-z analogues 
of local IR QSOs. On the other hand, targeted observations of 
high-z QSOs at submillimeter (submm) wavelength reveal a category 
of submm-loud QSOs (e.g., McMahon et al. 1999; Isaak et al. 2002). 
The average redshift and submm flux density of these submm detected QSOs 
by Stevens et al. (2005) are consistent with SMGs, but their median X-ray flux 
is 30 times higher than those of SMGs selected from blank surveys, indicating 
much higher $\Mdot$ than those of SMGs (see also Alexander et al. 2005b). 
Similarly, the average redshift and submm flux density of submm detected bright QSOs 
by Priddey et al. (2003a) are also comparable to those of X-ray selected QSOs 
by Stevens et al. (2005). Note also that most submm-loud QSOs by 
Isaak et al. (2002) have also been detected at $1.2$\,mm by Omont et al. (2001).

Given that several groups have investigated the properties of high-z 
bright QSO samples using targeted (sub)mm observations (e.g. Carilli et al. 2001; 
Omont et al. 2001, 2003; Priddey et al. 2003a,b; Stevens et al. 2005), 
it is natural to search from these samples 
the high-z analogues of the local IR QSOs, and study their physical 
properties. We are particularly interested in how the star formation and 
accretion processes are related to each other in the extreme environments and which of 
these dominates the heating of dust that gives rise to the rest-frame (thermal) 
submm emissions (e.g., Carilli et al. 2001; Isaak et al. 2002). 
To achieve this, we extend the technique developed 
in Paper I to the high-z (sub)mm observed QSO samples. 
This paper is structured as follows. In Sect.~2, we describe 
how the samples are compiled. In Sect.~3, 
we discuss how physical parameters are estimated. The main results 
are presented and discussed in Sect.~4. Finally, 
in Sect.~5, we briefly summarize our results. 
Throughout this paper we adopt a cosmology with 
a matter density parameter $\Omega_{\rm m}=0.3$, a cosmological constant 
$\Omega_{\rm \Lambda}=0.7$ and 
a Hubble constant of $H_{\rm 0}=70\,{\rm km \, s^{-1} Mpc^{-1}}$. 

\section{Samples}
\label{sect:Obs}

~~~~As discussed above, high-z QSO samples with 
submm or mm wavelength observations are needed for our purpose. 
First we collect several high-z optically selected QSO samples with mm ($1.2$\,mm) observations, taken from 
Carilli et al. (2001) and Omont et al. (2001, 2003). 
The $1.2$\,mm observations were made using the Max-Planck Millimeter 
Bolometer (MAMBO; Kreysa et al. 1998) 
on the IRAM 30m telescope on Pico Veleta in Spain. 
The details are listed below. Note that 
the $B$-band absolute magnitudes ($M_B$) in the following descriptions 
are taken directly from the cited papers, appropriate for a cosmology with 
$H_{\rm 0}=50\,{\rm km \, s^{-1} Mpc^{-1}}$, $\Omega_{\rm m}=1$ 
and $\Omega_\Lambda=0$. 

\begin{enumerate}
\item[(1)] A QSO sample taken from Carilli et al. (2001) consists of 41 sources, which were found in the Sloan Digital Sky Survey (SDSS). They span a range of 
$M_{\rm B}=-26.1$ to $-$28.8 and a redshift range from $z=3.6$ to 5.0. 
16 out of 41 objects have $1.2$\,mm flux densities 3 times greater 
than the rms noise (3$\sigma$).
\item[(2)] A QSO sample taken from Omont et al. (2001) consists of 65 objects, which 
were selected from Palomar Sky Survey (PSS). They have 
$M_{\rm B} < -27.0$ and $3.9 < z < 4.5$. 
21 out of these 65 objects were detected with flux densities greater than 3$\sigma$ at $1.2$\,mm. 
\item[(3)] An optically luminous ($M_{\rm B} < -27.0$) but radio quiet QSO sample 
taken from Omont et al. (2003) consists of 35 objects at redshift $1.8 < z < 2.8$. 
9 were detected at $1.2$\,mm with flux densities at levels $\geq 3\sigma$.
\end{enumerate}

As the Submillimeter Common User Bolometer Array (SCUBA) detector on the JCMT 
has similar capabilities with MAMBO, 
we also collect high-z optically selected QSO samples observed 
at submm ($850\mu{\rm m}$), taken from McMahon et al. (1999), 
Isaak et al. (2002) and Priddey et al. (2003a,b): 
\begin{enumerate}
\item[(1)] An optically luminous ($M_{\rm B} < -27$) and radio quiet QSO 
sample from MaMchon et al. (1999) with redshift $z \ga 4$. 
7 out of the 10 QSOs were detected at levels $\geq 3\sigma$ at $850\mu{\rm m}$.
\item[(2)] An optically luminous ($M_{\rm B} < -27.5$) and radio quiet QSO 
sample from Isaak et al. (2002) consists of 38 objects with redshift $z \geq 4$. 
8 were detected at $850\mu{\rm m}$ with flux densities at levels $\geq 3\sigma$.
\item[(3)] An optically luminous ($M_{\rm B} < -27.5$) and radio quiet QSO sample
taken from Priddey et al. (2003a) consists of 57 objects with redshift $1.5 < z < 3.0$. 
9 were detected at $850\mu{\rm m}$ with flux densities at levels $\geq 3\sigma$.
\item[(4)] An optically selected QSO sample with redshift $>4.9$ from 
Priddey et al. (2003b) consists of 14 objects, among which 4 were detected at 
$850\mu{\rm m}$ at $\geq 4\sigma$ levels (see Priddey et al. 2003b for detail).
\end{enumerate} 
In the following, we denote high-z sources as detected 
(non-detected) at $1.2$\,mm or at $850\mu{\rm m}$ if 
their flux densities are above (below) three times the rms noise level.

Besides these mm ($1.2$\,mm) and submm ($850\mu{\rm m}$) 
observed high-z QSOs samples, we 
also collect high-z QSOs with ultraluminous or hyperluminous FIR 
luminosities ($> 10^{12}$ or $10^{13} L_\odot$) detected by other means, 
to verify the assumption in our method  as many of them have unambiguous 
starbursts which dominate their infrared emissions (see \S4.1; 
Stevens et al. 2005; Carilli et al. 2005). Specifically, we include 
(1) 19 X-ray absorbed, Compton-thin QSOs with submm ($450 \mu{\rm m}$ and 
 $850\mu{\rm m}$) photometry (Stevens et al. 2005), among which 8 were 
detected at $850\mu{\rm m}$, and for comparison, 19 X-ray 
unabsorbed QSOs observed but non-detected at submm ($850 \mu{\rm m}$, 
Page et al. 2004). 
The submm photometry observations at $450 \mu{\rm m}$ and 
$850\mu{\rm m}$ for these QSOs were performed using SCUBA. 
The redshift range of this sample is $1<z<3$. 
Note that two of the X-ray absorbed, Compton thin QSOs treated as detections 
were observed with 2 $<S/N<$ 3 as indicated by Stevens et al. (2005). 
(2) two high-z QSOs (B1202$-$0725 and J1148+5251) with HCN and CO 
observations (Carilli et al. 2002; Walter et al. 2003; Isaak et al. 2004; 
Carilli et al. 2005) and one (B1335$-$0417) with only CO 
detections (Carilli et al. 2002). 
The $B$-band absolute luminosity and $1.25$\,mm 
flux densities for B1202$-$0725 and B1335$-$0417 are adopted from 
Omont et al. (1996); for J1148+5251, they are taken from Robson et al. 
(2004) and Bertoldi et al. (2003) respectively.

The comparison samples at low redshift are taken from paper I. 
Briefly, these include an IR QSO sample, an optically selected 
Palomar-Green QSO (PG QSO) sample, 
and a narrow-line Seyfert 1 galaxy (NLS1) sample. 
IR luminosities and bolometric luminosities are available for all these 
objects (see paper I for details).

As described above, most target QSOs were compiled from heterogeneous 
flux-limited samples, so they likely suffer from some selection biases. 
For example, QSOs at the bright end of the luminosity function (with 
high $\Mdot$) are favored in our samples. 
Nevertheless, the selection of high $\Mdot$ objects will not have significant 
impacts on our results as our main purpose is to compare the relative growth 
of black holes and spheroids (i.e., the ratio of $\Mstardot$ and $\Mdot$); 
we return to this point in Sect.~4.2. 

\section{Estimation of physical parameters}

~~~~For the local sample, the values of various physical parameters except the 
star formation rate (see \S3.2) were taken directly from paper I; 
we refer the readers to that paper for 
detail. Below we discuss how we derive the physical parameters for 
the high-z objects and the values are listed in Tables 1 and 
3 for mm and submm detected QSOs respectively.

\subsection{Accretion Rates}

~~~~The accretion rates are derived from the bolometric luminosities 
following the formula given in paper I: 
\begin{equation}
\Mdot=6.74\,\myear {L_{\rm bol} \over 10^{13}L_\odot}.
\label{mdoteq}
\end{equation} 
For the objects not selected from X-ray observations, 
the bolometric luminosities were estimated 
from the absolute $B$-band magnitude 
converted to our adopted cosmology. A bolometric correction factor 
of 9.74 was adopted according to Vestergaard (2004). For the X-ray selected 
QSOs, the bolometric luminosities are calculated from the X-ray 
luminosities, $L_X(0.5-2\,{\rm KeV})$, adopting 
a value of 33.3 as the bolometric correction factor 
(Stevens et al. 2005). The bolometric 
luminosities derived using two different methods are 
on average in agreement with each other. This can be seen from the 
application of these two methods to the X-ray absorbed QSOs that also 
have $B$-band magnitudes.

\subsection{Star Formation Rates}

~~~~The star formation rates are estimated from the 
monochromatic luminosities at $60\mu{\rm m}$ ($\L60=\nu L_\nu(60\mu{\rm m})$), 
\begin{equation}
\Mstardot \approx 3.26\, \myear {{L}_{60\mu{\rm m}}\over 10^{10}L_\odot}.
\label{eq:sfreq2}
\end{equation} 
This equation is obtained by the application of the conversion factors among 
the luminosities with different infrared wavelength coverage to the Kennicutt 
star formation rate law (Kennicutt 1998; see paper I for details). 
The star formation rate derived using equation (2) is 
$\sim 50\%$ of that using the formula given in paper I. This is because we find that 
the FIR luminosity between $40$ and $120\mu {\rm m}$ in our case (for local IR QSOs) is 
roughly equal to, rather than a factor of two of, $\L60$, 
as assumed in Paper I (see also Martin et al. 2005). 
This downward revision is also consistent with the prescription 
given by Rowan-Robinson (2000). Therefore, we recalculated $\Mstardot$ 
for IR QSOs using equation (2). This does not, however, affect our main conclusions in paper I (as the scatter around the mean relation is large).

For $1.2$\,mm detected high-z QSOs, we estimated the rest-frame monochromatic 
luminosity at $60\mu{\rm m}$ from the observed flux density at $1.2$\,mm 
by assuming that the rest-frame FIR spectral energy 
distribution (SED) can be described 
by a greybody spectrum with a dust temperature of 41\,K 
and a dust emissivity index ($\beta$) of 1.95. These parameters were derived by 
Priddey \& McMahon (2001) by fitting the photometric measurements at submm 
and mm wavebands for several high-z (z $ > 4$) quasars. 
Specifically, for QSOs with $1.2$\,mm observations, we apply the 
k-correction and obtain the rest-frame flux density at $1.2\,{\rm 
  mm}/(1+z)$; this step fixes the overall normalization of the greybody 
SED template, which can then be used to derive the rest-frame flux density at 
$60\mu{\rm m}$. The star formation rate is then 
obtained using equation (2). For the 
X-ray selected and other $850\mu {\rm m}$ detected QSOs, the same method 
is applied at $850\mu{\rm m}$ instead of $1.2$\,mm.

In previous studies, an assumption of a rest-frame FIR SED with a dust 
temperature 
of $\sim 50$\,K and a dust emissivity index of $\beta \sim 1.5$ is often used 
(e.g. Omont et al. 2001; Carilli et al. 2001). Recent studies of 
the SEDs of more than ten high-z QSOs 
give a best-fit of temperature of $52\pm3$\,K and a dust emissivity index of 
$\beta=1.5\pm0.1$, with individual dust temperatures ranging from 40 to 60K 
(Beelen et al. 2006). If we adopt these dust parameter values, rather than the 
Priddey \& McMahon (2001) SED fitting results, our estimates of 
$\L60$ are changed by only $\sim 20-30\%$. However, when the individual dust 
temperature differences (40-60K) are concerned, the estimates will be 
changed by more than 0.5 dex for $\beta=1.5\pm0.1$. 

It is worth noting that the prescription we used here to obtain the 
star formation rate differs from that adopted by 
Carilli et al. (2001) and Omont et al. (2001, 2003), 
which gives a star formation rate that is several times 
lower than that from eq. (2). The differences arise 
from different methods in deriving the IR or FIR luminosity, 
and different assumptions about the initial mass function, stellar evolutionary theory 
and the wavebands in which the absorbed starburst's total luminosity is re-emitted. 
As discussed by Kennicutt (1998), the star formation rates derived with 
different methods depend on a number of assumptions and  are not 
precise; each method offers a useful means of estimating the star formation 
activity. For consistency, we use eq. (2) to estimate the 
$\Mstardot$ for objects in the local universe and at high redshift.

\section{Results and discussions}

\subsection{(Sub)mm-Quiet and (Sub)mm-Loud QSOs}

~~~~Fig.~1a
shows the rest-frame monochromatic luminosity at $60\mu{\rm m}$, $\L60$, versus 
the bolometric luminosity associated with the central AGNs 
for all low-z sources and the $1.2$\,mm observed high-z QSOs. 
The regression line in Fig.~1a
is the best fit line for 
low-z typical type 1 AGNs (PG QSOs and NLS1s). The 
close correlation between $\L60$ and $\Lbol$ for low-z 
typical type 1 AGNs suggests 
that their FIR emissions are mainly 
powered by the central AGNs (see paper I and Haas et al. 2003). 
It is clear from Fig.~1a that all 
the high-z QSOs detected at $1.2$\,mm 
are located above the regression line just as the low-z IR QSOs.

The majority ($\sim$ 70\%) of optically selected QSOs are, however, not detected at 
$1.2$\,mm above the $3\sigma$ level (e.g. Carilli et al. 2004; Momjian 
et al. 2005). We use the stacking method (see Stevens et al. 2005) to obtain 
the mean value of the rest-frame $\L60$ for the 95 QSOs not detected at 
$1.2$\,mm. For each sample (see Sect.~2), we divide 
the redshift into bins with width of  $\approx 0.5$. The bolometric luminosity is 
simply the average of the objects in each bin. The $1.2$\,mm  flux density is obtained as the mean 
of the sources in each bin weighted by their uncertainties. 
The rest-frame $\L60$ is calculated in the way as described in Sect.~3.2, 
with the obtained mean flux densities and the mean bin redshift. 
The error bars of the binned data are calculated using error propagation 
by weighting the uncertainties of the observed flux densities at $1.2$\,mm. 
The results from stacking are listed in Table 2 and plotted in Fig.~1a. It is 
striking that these data points are all around the regression line 
inferred from the low-z typical type 1 AGNs.

A similar analysis was performed for high-z QSOs observed at 
$850\mu{\rm m}$ (instead of 1.2mm). These include QSOs selected optically and 
from 
X-ray (see Sect. 2). Fig.~1b shows $\L60$ 
vs. $\Lbol$ for all low-z type 1 AGNs, IR QSOs, 
high-z submm detected QSOs and submm non-detected QSOs. 
Fig.~1b clearly shows that 
all $850\mu{\rm m}$ non-detected QSOs except the X-ray absorbed ones are around 
the regression line derived from low-z type 1 AGNs. 
In contrast, all the $850\mu{\rm m}$ detected QSOs are above this regression line.

The similar behaviours of high-z QSOs observed at $850\mu{\rm m}$ by 
SCUBA and those observed at $1.2$\,mm by MAMBO 
in the relation of FIR luminosities to bolometric luminosities show that 
the difference between (sub)mm detected and non-detected QSOs may be 
real, rather than from the effects of instrument sensitivities. 
In fact, the samples used here were observed with different sensitivities, even for measurements taken by 
the same instrument. For $1.2$\,mm observations by MAMBO, the typical rms 
sensitivities vary from 0.5 mJy to 1.4mJy; while for $850\mu{\rm m}$ observations 
by SCUBA, the sensitivities of the surveys change from 1.5 mJy to 3.3 mJy.
In a word, the sensitivities vary by a factor of 2-3 
for different observations (e.g. Omont et al. 2001, 2003; Carilli et al. 2001; 
MaMchon et al. 1999; Issak et al. 2002). 
Observations with different sensitivities lead to 
significantly different weighted mean flux densities for non-detected 
QSOs shown in Tables 2 and 4. 
Nevertheless, the positions of the (sub)mm non-detected QSOs 
are not strongly influenced by the sensitivities as they 
are all located around the regression line inferred for low-z typical 
QSOs. This suggests that the high-z (sub)mm non-detected QSOs may be the 
analogues of typical QSOs seen locally and the 
FIR emission from these high-z objects are powered by 
AGNs, just as the low-z typical QSOs. Of course, the high-z QSOs have higher 
accretion rates than their local counterparts.

On the other hand, for high-z QSOs detected at $1.2$\,mm and $850\mu{\rm m}$, 
their rest-frame $\L60$ 
are all above the regression line for typical QSOs, implying that these 
objects are the analogues of low-z IR QSOs. 
For both the local and high-z samples, the excess 
FIR emission relative to the regression lines is probably provided by an additional 
energy source, namely massive starbursts in these objects. 

Our conclusions are supported by the CO and/or HCN observations for three 
additional high-z QSOs (B1335$-$0417, 
B1202$-$0725 and J1148$+$5251, shown as crosses in Fig.~1a and
Fig.~1b), 
which revealed massive molecular gas reservoirs in these three QSOs 
($10^{10}$ to $10^{11}$ \msun, e.g., Carilli et al. 2002, 2004; Walter et al. 
2003, 2004). 
Therefore, massive starbursts are occurring at the host spheroids 
of these objects and provide the dominant energy source which heats the dust. 
In Fig.~1a and Fig.~1b, these three QSOs are clearly 
located above the regression line for typical QSOs and they 
mix well with all the (sub)mm detected QSOs. 
In addition, two $1.2$\,mm detected QSOs (J140955.5$+$562827, Omont 2003, and PSS J2322$+$1944, 
Omont 2001, shown as crossed squares
in Fig.~1a and Fig.~1b) have also been detected by CO observations. 
Note that only 1 (PSS J2322$+$1944) out of these 5 QSOs are magnified by 
gravitational lensing. In any case, lensing magnification should not be 
statistically important for most of our objects according to Vestergaard (2004).

In fact, using deep radio observations, it has already been suggested that the 
physics of submm-loud QSOs and submm-quiet QSOs may be different 
(Petric \& Carilli 2004). 
Furthermore, the extended dust emission regions (larger than 1 kpc) 
of (sub)mm-loud QSOs rule out the central AGN heating model 
(e.g., Momjian et al. 2005 and references therein). Stevens et al. (2005) also 
argued that 
the submm emission of submm detected, X-ray absorbed QSOs is attributed to dust 
heated by hot young stars. The approach we adopt here is different but 
we reach the same conclusion -- 
the high-z (sub)mm-loud QSOs are low-z analogues of IR QSOs and their 
ultraluminous/hyperluminous FIR emissions are mainly from dust heated by massive 
starbursts. As we show below, they are at a transition phase with 
rapid growth of black holes and their host spheroids.

\subsection{Coeval Growth of Black Holes and Host Spheroids}

~~~~Given that the ultraluminous/hyperluminous FIR emissions are 
mainly from dust heated by massive 
starbursts for both low-z IR QSOs and high-z (sub)mm-loud QSOs, 
we can estimate the star formation rates for these high-z 
(sub)mm-loud QSOs using the same method as that for low-z IR QSOs. 
The star formation rates are estimated from $\L60$, 
after subtracting the contribution from the central AGNs, 
namely assuming the AGN contribution to $\L60$ 
for (sub)mm-loud QSOs follows the same regression line as typical QSOs shown 
as the solid line in Fig.~1.

Fig.~2a and b shows the star formation rate versus 
the accretion rate for IR QSOs and high-z (sub)mm-loud QSOs. 
It is obvious from Fig.~2
that for QSOs with larger accretion rates and hence higher bolometric luminosities, 
their star formation rates are also higher. This trend directly indicates 
that the massive galaxies build their spheroid stellar masses and central 
black holes quicker than less massive ones, and hence the most massive 
galaxies host the most luminous QSOs in their centers. 
Although this trend has been noticed in paper I, 
the combination of high-z (sub)mm-loud QSOs with the local 
IR QSOs provides a much larger dynamical range for studying the 
relation between $\Mstardot$ and $\Mdot$ than 
either sample alone. This is appropriate because the underlying 
physics of QSOs at low and high redshift may be similar (Fan et al. 2004; 
Vestergaard 2004 and references therein; see also below). Thus we 
can extend the dynamic range in the bolometric luminosity by 
combining low- and high-redshift samples together. 
In contrast, this trend was not found based on 
high-z (sub)mm observed QSOs alone (e.g. Isaak et al. 2002; Omont et al. 2003) 
because of limited dynamical range and large scatters involved. 
In addition, the inclusion of both (sub)mm-loud and (sub)mm-quiet QSOs in 
previous analyses also obscures the trend as these 
two types of objects may be at physically different 
evolutionary stages (see below).

As can be seen from Fig.~2a and b, for a fixed $\Mdot$, 
there is roughly an order of magnitude scatter in $\Mstardot$. The large 
scatters may be partly because of the calibration errors in equation (2),
and the differences in the FIR SED in QSOs. 
Nevertheless, it is quite likely that $\Mstardot/\Mdot$ does vary from 
object to object. Therefore 
it is worth investigating how the ${\Mstardot/\Mdot}$ changes with redshifts and bolometric luminosities of QSOs. 

Fig.~3a and b shows ${\Mstardot/\Mdot}$ versus 
redshift for low-z IR QSOs and high-z (sub)mm-loud QSOs. The sizes of the 
data points are scaled by their accretion rates, i.e., 
by their bolometric luminosities. 
It is obvious from Fig.~3a and b that 
there is an absence of QSOs with large ${\Mstardot/\Mdot}$ at higher redshift. 
This may be caused by the Malmquist bias -- for a flux-limited sample, as 
the redshift (distance) increases, more luminous objects are 
preferentially selected. 
In our case, the objects with lower accretion rates at high-z could have been missed, 
although the samples we used here are not strictly flux-limited samples 
(most of our QSOs are optically luminous QSOs with $M_{\rm B}< -27.0$). 
To illustrate the influence of Malmquist bias quantitatively, we plot three dotted 
curves in Fig.~3a and b corresponding to 
a bolometric luminosity of $10^{11}L_\odot$, $10^{12}L_\odot$ 
and $10^{13}L_\odot$ at redshift 2 respectively, which cover the whole 
range of the bolometric luminosity of low-z IR QSOs. For each curve, 
we assume the relation between the star formation rate and the accretion rate 
by simply fitting a regression line to the data points in Fig.~2a 
using survival analysis (Isobe et al. 1986). In an optically selected 
flux-limited sample, the objects above these dotted lines will be missed during 
the observations with flux limits assumed above. From the shapes 
of these dotted lines, we can see that the 
higher ratios at lower redshift can be reproduced by the Malmquist bias, 
so the lack of high $\Mstardot/\Mdot$ QSOs at high z may not be real.

There is a well-studied high-z ($z=5.5$) optically faint 
($M_{\rm B} \sim -24.2$) QSO -- RDJ030117$+$002025 (e.g., Stern et al. 2000; 
Bertoldi \& Cox 2002; Staguhn et al. 2005) to test the above argument. 
The pentagram in all three figures 
represents this object.  From the location of RDJ030117$+$002025 
in Figs.~1-3, we can see that its behavior is similar to the low-z IR QSOs. 
This suggests that at least some distant optically faint QSOs have similar 
properties to the low-z IR QSOs, which also verifies the 
combination of 
low-z IR QSOs and high-z (sub)mm-loud QSOs in Fig.~2.
However, more (sub)mm observations of high-z optically faint QSOs 
are needed to firmly establish this.

On the other hand, it can be seen from Fig.~3a
and b that 
the ${\Mstardot/\Mdot}$ of the low-z IR QSOs and high-z 
(sub)mm-loud QSOs both have broad ranges and 
there is a clear trend that as the bolometric luminosity (indicated by the 
size of data points) increases, the ${\Mstardot/\Mdot}$ decreases. 
Thus the behavior of the relative growth of host spheroids and their central 
black holes is correlated with the power of QSOs -- 
the optically more luminous QSOs correspond to a phase with higher 
accretion rates and relatively low star formation rates, which results 
in the absence of optically luminous QSOs (shown as symbols with larger sizes 
in Fig.~3a and b) with large ${\Mstardot/\Mdot}$. 
This trend is consistent with the simulation prediction of 
Cattaneo et al. (2005, their Fig 11). 

In Fig.~3a and b we also indicate the positions of 
SMGs (shown as pentagon) using their data in the Chandra Deep Field North 
with deep Keck spectroscopic data in Alexander et al. (2005b). 
The characteristic position of SMGs is obtained 
using the mean redshift and the mean ${\Mstardot/\Mdot}$ 
in the Alexander et al. (2005b) sample. 
It is clear from Fig.~3a and b that the SMGs are located 
above the FIR ultraluminous/hyperluminous QSOs. 

The locations of SMGs and FIR ultraluminous/hyperluminous QSOs 
in Fig.~3 suggest a possible evolutionary picture 
in terms of ${\Mstardot/\Mdot}$, from top to bottom. 
SMGs represent prodigious 
starbursts triggered by interactions and major mergers as revealed by Hubble 
Space Telescope images in the rest-frame 
ultra-violet wavebands (Conselice et al. 2003; Smail et al. 2004). Although 
$> 38^{+12}_{-10}\%$ SMGs host AGNs, 
the dominant rest-frame FIR energy output for 
most of them is still from starbursts (Alexander et al. 2005b) and this 
population is expected to be at a pre-QSO phase, during which the central 
black holes grow more slowly compared with their spheroids. As the gas 
falls toward the centers of galaxies, the accretion rate 
increases while the massive starbursts continue, experiencing a phase of 
simultaneous growth of central black holes 
and their spheroids. This coeval phase may be quite short 
($10^{7}$ to $10^{8}$ years) if the (sub)mm-detected 
QSO fraction (20-30\%) can be interpreted as the relative duty cycle 
(e.g. Carilli et al. 2004). 
The very high star formation rates and accretion rates 
in this phase lead to a rapid increase of the masses of the 
central black holes and spheroids. 
As the black hole mass increases, the central AGN becomes more energetic, and the AGN feedback 
process may heat up and blow away the surrounding 
gas and dust, leading to a decrease in the star formation rate and ${\Mstardot/\Mdot}$. 
Eventually, it enters the typical QSO phase with high accretion rate but 
low star formation rate. The QSOs at this stage will be located 
at the bottom of Fig.~3.

The picture described above corresponds well with 
the evolutionary sequence for low-z gas-rich major mergers, 
from ultraluminous IR galaxies  to IR QSOs (or FIR luminous Seyfert 1s; 
Sanders et al. 1988a,b), and finally to $L^\ast$ 
elliptical galaxies. The difference is that high-z SMGs have more gas 
and may evolve into giant elliptical galaxies. While not every single object 
may fit in this picture, e.g., the high-z QSOs with large black hole mass
and relatively low host dynamical mass may have a different evolutionary 
path, many ultraluminous IR galaxies (at low redshift) and SMGs 
(at high redshift) will evolve 
into $L^\ast$ and giant ellipticals respectively along this path.

\section{Summary}

~~~~In this paper we have examined the properties of high-z 
QSOs samples observed at $1.2$\,mm, $850\mu{\rm m}$ or with 
CO and/or HCN observations. The redshift of these objects spans from redshift 
1 to 6.42. 
Applying the same criteria established for local 
type 1 AGNs (Hao et al. 2005), we found that statistically, the (sub)mm 
detected and non-detected high-z QSOs are analogues of local IR QSOs and 
typical type 1 AGNs, respectively. 
We postulated that the underlying physics of (sub)mm detected and non-detected 
high-z QSOs may be different, and they correspond to different 
phases in the interplay between AGNs and spheroids formation.

For (sub)mm-loud QSOs, the FIR ultraluminous/hyperluminous emissions 
are from dust heated predominantly by starbursts. 
By combining low-z IR QSOs and high-z (sub)mm-loud 
QSOs, we found a clear trend that the higher the accretion rate, the larger 
the star formation rate. It directly indicates that 
the most massive galaxies host the most luminous QSOs at their centers. 
We also found that the relative growth of black holes and their 
host spheroids depends on the intensity of QSO activities, 
in qualitative agreement with theoretical expectations. 
We also compared the properties of SMGs, ultraluminous or hyperluminous FIR 
QSOs and typical QSOs, and constructed a possible 
evolution scenario among these objects. Clearly substantial growth of 
black holes and their host spheroids can occur in these objects. 
Future studies of these objects are therefore important for 
understanding how formation of spheroids and AGNs are inter-connected and 
how the $\Mbh$-$M_\star$ relation arises.

\begin{acknowledgements}
We thank P. Cox, X. Fan and Y. Gao for advice and helpful discussions. Thanks are also 
due to R. Kennicutt for helpful discussions on the star formation rate 
estimators. This project is supported by the NSF of China 10333060 and 10778622. 
SM acknowledges partial travel support from the 
Chinese Academy of Sciences and a visiting professorship from Tianjin Normal 
University.
\end{acknowledgements}

\clearpage
\begin{table}[]
\scriptsize
  \caption[]{Various Physical Parameters For $1.2$\,mm Detected QSOs } 
  \begin{tabular}{lccccccc}
  \hline\noalign{\smallskip}
Name &  Redshift  & S$_{\rm 1.2mm}$ & log$(\frac{L_{\rm bol}}{L_{\odot}})$ & log$(\frac{L_{60\mu{\rm m}}}{L_{\odot}})$ & log$(\frac{\Mdot}{\myear})$ & log$(\frac{\Mstardot}{\myear})$ & log$(\frac{\Mstardot}{\Mdot})$                    \\
(1) & (2) & (3) & (4) & (5) & (6) & (7) & (8) \\
\hline\noalign{\smallskip}
 & & & & Carilli et al. (2001) & & &   \\
\noalign{\smallskip}
  \hline\noalign{\smallskip}
J012403.78$+$004432.7 & 3.81 & 2.0$\pm$0.3\phn & 14.272 & 12.884 & 2.101 & 3.089 & 0.988 \\
J015048.83$+$004126.2 & 3.67 & 2.2$\pm$0.4\phn & 14.094 & 12.939 & 1.923 & 3.283 & 1.360 \\
J023231.40$-$000010.7 & 3.81 & 1.8$\pm$0.3\phn & 13.612 & 12.838 & 1.441 & 3.271 & 1.830 \\
J025112.44$-$005208.2 & 3.78 & 2.4$\pm$0.6\phn & 13.691 & 12.966 & 1.520 & 3.411 & 1.891 \\
J025518.58$+$004847.6 & 3.97 & 2.1$\pm$0.4\phn & 14.066 & 12.890 & 1.895 & 3.220 & 1.325 \\
J032608.12$-$003340.2 & 4.16 & 1.5$\pm$0.4\phn & 13.885 & 12.726 & 1.714 & 3.045 & 1.331 \\
J033829.31$+$002156.3 & 5.00 & 3.7$\pm$0.3\phn & 13.634 & 13.050 & 1.463 & 3.514 & 2.051 \\
J111246.30$+$004957.5 & 3.92 & 2.7$\pm$0.5\phn & 13.997 & 13.004 & 1.826 & 3.402 & 1.576 \\
J122600.68$+$005923.6 & 4.25 & 1.4$\pm$0.4\phn & 13.946 & 12.688 & 1.775 & 2.950 & 1.175 \\
J123503.04$-$000331.8 & 4.69 & 1.6$\pm$0.4\phn & 13.523 & 12.709 & 1.352 & 3.129 & 1.777 \\
J141205.78$-$010152.6 & 3.73 & 4.5$\pm$0.7\phn & 13.759 & 13.244 & 1.588 & 3.718 & 2.130 \\
J141332.35$-$004909.7 & 4.14 & 2.5$\pm$0.5\phn & 13.840 & 12.950 & 1.669 & 3.367 & 1.698 \\
J142647.82$+$002740.4 & 3.69 & 3.9$\pm$0.8\phn & 13.670 & 13.186 & 1.499 & 3.661 & 2.162 \\
J144758.46$-$005055.4 & 3.80 & 5.4$\pm$0.8\phn & 13.644 & 13.316 & 1.473 & 3.803 & 2.330 \\
J161926.87$-$011825.2 & 3.84 & 2.3$\pm$0.6\phn & 13.612 & 12.942 & 1.441 & 3.393 & 1.952 \\
J235718.35$+$004350.4 & 4.34 & 1.8$\pm$0.6\phn & 13.639 & 12.789 & 1.468 & 3.206 & 1.738 \\
\hline\noalign{\smallskip}
 & & & & Omont et al. (2001) & & & \\
\noalign{\smallskip}\hline
PSSJ0209$+$0517 & 4.18 & 3.3$\pm$0.6\phn & 14.241 & 13.067 & 2.070 & 3.416 & 1.346 \\
PSSJ0439$-$0207 & 4.40 & 2.3$\pm$0.7\phn & 14.003 & 12.891 & 1.832 & 3.246 & 1.414 \\
PSSJ0808$+$5215 & 4.44 & 6.6$\pm$0.6\phn & 14.484 & 13.345 & 2.313 & 3.728 & 1.415 \\
PSSJ1048$+$4407 & 4.40 & 4.6$\pm$0.4\phn & 13.963 & 13.192 & 1.792 & 3.639 & 1.847 \\
PSSJ1057$+$4555 & 4.12 & 4.9$\pm$0.7\phn & 14.520 & 13.244 & 2.349 & 3.571 & 1.222 \\
BRB1117$-$1329 & 3.96 & 4.1$\pm$0.7\phn & 14.278 & 13.181 & 2.107 & 3.564 & 1.457 \\
BRB1144$-$0723 & 4.15 & 6.0$\pm$0.7\phn & 14.080 & 13.329 & 1.909 & 3.783 & 1.874 \\
PSSJ1226$+$0950 & 4.34 & 2.8$\pm$0.7\phn & 13.963 & 12.981 & 1.792 & 3.380 & 1.588 \\
PSSJ1248$+$3110 & 4.32 & 6.3$\pm$0.8\phn & 14.043 & 13.335 & 1.872 & 3.794 & 1.922 \\
PSSJ1253$-$0228 & 4.00 & 5.5$\pm$0.8\phn & 13.878 & 13.305 & 1.707 & 3.775 & 2.068 \\
PSSJ1317$+$3531 & 4.36 & 3.7$\pm$1.1\phn & 14.003 & 13.100 & 1.832 & 3.522 & 1.690 \\
PSSJ1347$+$4956 & 4.56 & 5.7$\pm$0.7\phn & 14.325 & 13.271 & 2.154 & 3.671 & 1.517 \\
PSSJ1403$+$4126 & 3.85 & 1.5$\pm$0.5\phn & 13.716 & 12.755 & 1.545 & 3.145 & 1.600 \\
PSSJ1418$+$4449 & 4.32 & 6.3$\pm$0.7\phn & 14.443 & 13.335 & 2.272 & 3.726 & 1.454 \\
PSSJ1535$+$2943 & 3.99 & 1.9$\pm$0.6\phn & 13.838 & 12.844 & 1.667 & 3.231 & 1.564 \\
PSSJ1554$+$1835 & 3.99 & 6.7$\pm$1.1\phn & 13.638 & 13.392 & 1.467 & 3.883 & 2.416 \\
PSSJ1555$+$2003 & 4.22 & 3.1$\pm$0.6\phn & 13.961 & 13.036 & 1.790 & 3.451 & 1.661 \\
PSSJ1646$+$5514 & 4.04 & 4.6$\pm$1.5\phn & 14.479 & 13.224 & 2.308 & 3.557 & 1.249 \\
PSSJ1745$+$6846 & 4.13 & 2.5$\pm$0.7\phn & 13.800 & 12.951 & 1.629 & 3.376 & 1.747 \\
PSSJ1802$+$5616 & 4.16 & 2.8$\pm$0.9\phn & 13.761 & 12.997 & 1.590 & 3.438 & 1.848 \\
PSSJ2322$+$1944 & 4.11 & 9.6$\pm$0.5\phn & 14.240 & 13.537 & 2.069 & 4.001 & 1.932 \\
\hline\noalign{\smallskip}
 & & & & Omont et al. (2003) & & & \\
\noalign{\smallskip}\hline
KUV08086$+$4037 & 1.78 & 4.3$\pm$0.8\phn & 13.739 & 13.443 & 1.568 & 3.933 & 2.365 \\
093750.9$+$730206$^{a}$ & 2.52 & 3.8$\pm$0.9\phn & 14.368 & 13.304 & 2.197 & 3.704 & 1.507 \\
HS1002$+$4400 & 2.08 & 4.2$\pm$0.8\phn & 14.273 & 13.400 & 2.102 & 3.840 & 1.738 \\
HS1049$+$4033 & 2.15 & 3.2$\pm$0.7\phn & 14.235 & 13.273 & 2.064 & 3.693 & 1.629 \\
110610.8$+$640008$^{a}$ & 2.19 & 3.9$\pm$1.1\phn & 14.677 & 13.355 & 2.506 & 3.674 & 1.168 \\
140955.5$+$562827$^{a}$ & 2.56 & 10.7$\pm$0.6\phn & 14.329 & 13.749 & 2.158 & 4.227 & 2.069 \\
154359.3$+$535903$^{a}$ & 2.37 & 3.8$\pm$1.1\phn & 14.283 & 13.322 & 2.112 & 3.744 & 1.632 \\
HS1611$+$4719 & 2.35 & 4.6$\pm$0.7\phn & 14.043 & 13.408 & 1.872 & 3.875 & 2.003 \\
164914.9$+$530316$^{b}$ & 2.26 & 4.6$\pm$0.8\phn & 14.240 & 13.418 & 2.069 & 3.866 & 1.797 \\
\hline\noalign{\smallskip}
 & & & & high-z optically faint QSO & & & \\
\noalign{\smallskip}\hline
\noalign{\smallskip}
030117$+$002025$^{c}$ & 5.5 & 0.87$\pm$0.20\phn & 12.374 & 12.387 & 0.216 & 2.878 & 2.662 \\
\noalign{\smallskip}
\hline\noalign{\smallskip}
 & & & & High redshift QSOs with HCN/CO observations & & & \\
\noalign{\smallskip}\hline
BRI1202$-$0725 & 4.693 & 12.59$\pm$2.28\phn & 14.407 & 13.657 & 2.236 & 4.120 & 1.884 \\
BRI1335$-$0417 & 4.407 & 10.26$\pm$1.04\phn & 13.924 & 13.592 & 1.753 & 4.082 & 2.329 \\
J1148$+$5251 & 6.419 & 5.0$\pm$0.6\phn & 14.388 & 13.095 & 2.217 & 3.396 & 1.179 \\
\noalign{\smallskip}\hline
  \end{tabular}
\scriptsize 
Notes: Columns: (1) name. (2) redshift. (3) the observed flux 
density at $1.2$\,mm. (4) bolometric luminosity of 
AGN. (5) monochromatic luminosity at 60$\mu$m ($\nu L_\nu$) 
(6) accretion rate of central supermassive black hole in $\myear$. 
(7) star formation rate in $\myear$. 
(8) the ratio of the star formation rate to the accretion rate. 
The objects from different samples are separated and labeled.\\ 
$^{a}$The name with prefix $[{\rm VV96}]$J.\\ 
$^{b}$The name with prefix $[{\rm VV2000}]$J.\\ 
$^{c}$The name with prefix RDJ.
\end{table}

\clearpage
\begin{table}[]
\caption[]{Binned data for $1.2$mm non-detected QSOs}
\begin{center}
\begin{tabular}{ccccc}
\hline
\noalign{\smallskip}
\hline
Number & mean redshift & weighted mean S$_{\rm 1.2mm}$ & weighted mean log$(\frac{L_{\rm bol}}{L_{\odot}})$ & weighted mean log$(\frac {L_{60\mu{\rm m}}}{L_{\odot}})$ \\
\hline\noalign{\smallskip}
 & & &  Carilli et al. (2001) & \\
\noalign{\smallskip}\hline
\noalign{\smallskip}
 9 & 3.749 & 0.190$\pm$0.146\phn & 13.915 & 11.868$\pm$0.333\phn \\
  7 & 4.244 & 0.092$\pm$0.146\phn & 13.860 & 11.506$\pm$0.689\phn \\
  9 & 4.744 & 0.187$\pm$0.125\phn & 13.821 & 11.773$\pm$0.290\phn \\
\noalign{\smallskip}\hline
\noalign{\smallskip}
 & & & Omont et al. (2001)$^{a}$ & \\
\noalign{\smallskip}\hline
\noalign{\smallskip}
 44 & 4.232 & 0.385$\pm$0.128\phn &14.252 & 12.129$\pm$0.144\phn \\
\noalign{\smallskip}\hline
\noalign{\smallskip}
 & & & Omont et al. (2003) & \\
\noalign{\smallskip}\hline
\noalign{\smallskip}
 19 & 2.172 & 0.672$\pm$0.210\phn & 14.219 & 12.593$\pm$0.136\phn \\
  7 & 2.529 & 0.738$\pm$0.287\phn & 14.074 & 12.592$\pm$0.169\phn \\
\noalign{\smallskip}\hline
\noalign{\smallskip}
\end{tabular}
\end{center}
Notes: The binned data for the $1.2$mm non-detected QSOs in 
Carilli et al. (2001) and Omont et al. (2001, 2003). Columns: (1) 
the number of objects in each redshift bin with a bin width of 0.5. 
(2) the mean redshift in each bin. (3) the weighted mean flux density 
at $1.2$\,mm. (4) the average bolometric 
luminosity in each bin. (5) the weighted mean of the monochromatic 
luminosity at $60\mu{\rm m}$. \\ 
$^{a}$The redshift bin width is 0.56, which is the whole 
redshift range of this sample.
\end{table}
\label{lastpage}
	
\clearpage
\begin{table}[]
\scriptsize
\caption{Various Physical Parameters for $850\mu{\rm m}$ detected QSOs}
\begin{tabular}{lccccccc}
\hline\hline
Name & Redshift & S$_{\rm 850\mu m}$ & log$(\frac{L_{\rm bol}}{L_{\odot}})$ & 
log$(\frac{L_{60\mu{\rm m}}}{L_{\odot}})$ & 
log$(\frac{\Mdot}{\myear})$ & 
log$(\frac{\Mstardot}{\myear})$ & 
log$(\frac{\Mstardot}{\Mdot})$  \\ 
(1) & (2) & (3) & (4) & (5) & (6) & (7) & (8) \\
\hline\noalign{\smallskip}
 & & & & X-ray Absorbed QSOs$^{a}$ & & & \\
\noalign{\smallskip}\hline
005734.78$-$272827.4 & 2.19 & 11.7$\pm$1.2\phn & 12.796 & 13.339 & 0.625 & 3.847 & 3.222 \\
094144.51$+$385434.8 & 1.82 & 13.4$\pm$1.5\phn & 12.774 & 13.426 & 0.603 & 3.935 & 3.332 \\
094356.53$+$164244.1 & 1.92 & 3.0$\pm$1.2\phn & 13.296 & 12.769 & 1.125 & 3.231 & 2.106 \\
110431.75$+$355208.5 & 1.63 & 2.4$\pm$1.2\phn & 12.917 & 12.691 & 0.746 & 3.174 & 2.428 \\
110742.05$+$723236.0 & 2.10 & 10.4$\pm$1.2\phn & 13.475 & 13.295 & 1.304 & 3.788 & 2.484 \\
121803.82$+$470854.6 & 1.74 & 6.8$\pm$1.2\phn & 12.625 & 13.137 & 0.454 & 3.644 & 3.190 \\
124913.86$-$055906.2 & 2.21 & 7.2$\pm$1.4\phn & 13.037 & 13.126 & 0.866 & 3.626 & 2.760 \\
163308.57$+$570258.7 & 2.80 & 5.9$\pm$1.1\phn & 13.181 & 12.992 & 1.010 & 3.481 & 2.471 \\
\noalign{\smallskip}\hline
\noalign{\smallskip}
 & & & & Isaak et al. (2002) & & & \\
\noalign{\smallskip}\hline
\noalign{\smallskip}
PSSJ0452$+$0355 & 4.38 & 10.6$\pm$2.1\phn & 14.043 & 13.147 & 1.872 & 3.573 & 1.701 \\
PSSJ0808$+$5215 & 4.44 & 17.4$\pm$2.8\phn & 14.484 & 13.360 & 2.313 & 3.748 & 1.435 \\
PSSJ1048$+$4407 & 4.40 & 12.0$\pm$2.2\phn & 13.963 & 13.200 & 1.792 & 3.648 & 1.856 \\
PSSJ1057$+$4555 & 4.12 & 19.2$\pm$2.8\phn & 14.760 & 13.417 & 2.589 & 3.734 & 1.145 \\
PSSJ1248$+$3110 & 4.32 & 12.7$\pm$3.4\phn & 14.043 & 13.228 & 1.872 & 3.670 & 1.798 \\
PSSJ1418$+$4449 & 4.32 & 10.4$\pm$2.3\phn & 14.443 & 13.141 & 2.272 & 3.444 & 1.172 \\
PSSJ1646$+$5514 & 4.04 & 9.5$\pm$2.5\phn & 14.479 & 13.116 & 2.308 & 3.381 & 1.073 \\
PSSJ2322$+$1944 & 4.11 & 22.5$\pm$2.5\phn & 14.240 & 13.487 & 2.069 & 3.945 & 1.876 \\
\noalign{\smallskip}\hline
\noalign{\smallskip}
 & & & & McMahon et al. (1999) & & & \\
\noalign{\smallskip}\hline
\noalign{\smallskip}
BR2237$-$0607 & 4.55 & 5.0$\pm$1.1\phn & 14.245 & 12.813 & 2.074 & 2.960 & 0.886 \\
BRI0952$-$0115 & 4.43 & 14$\pm$2\phn & 14.084 & 13.266 & 1.913 & 3.709 & 1.796 \\
BR1033$-$0327 & 4.50 & 7$\pm$2\phn & 14.045 & 12.962 & 1.874 & 3.332 & 1.458 \\
BR1117$-$1329 & 3.96 & 13$\pm$1\phn & 14.238 & 13.257 & 2.067 & 3.672 & 1.605 \\BR1144$-$0723 & 4.14 & 7$\pm$2\phn & 14.000 & 12.978 & 1.829 & 3.367 & 1.538 \\
BR1202$-$0725 & 4.69 & 42$\pm$2\phn & 14.407 & 13.732 & 2.236 & 4.203 & 1.967 \\BRI1335$-$0417 & 4.40 & 14$\pm$1\phn & 13.923 & 13.267 & 1.752 & 3.729 & 1.977 \\
\noalign{\smallskip}\hline
\noalign{\smallskip}
 & & & & Priddey et al. (2003a) z$>\sim$2 & & & \\
\noalign{\smallskip}\hline
\noalign{\smallskip}
LBQSB0018$-$0220 & 2.56 & 17.2$\pm$2.9\phn & 14.289 & 13.476 & 2.118 & 3.927 & 1.809 \\
HSB0035$+$4405 & 2.71 & 9.4$\pm$2.8\phn & 14.334 & 13.201 & 2.163 & 3.575 & 1.412 \\
HSB0211$+$1858 & 2.47 & 7.1$\pm$2.1\phn & 14.167 & 13.099 & 1.996 & 3.484 & 1.488 \\
HSB0810$+$2554 & 1.50 & 7.6$\pm$1.8\phn & 14.562 & 13.198 & 2.391 & 3.475 & 1.084 \\
HSB0943$+$3155 & 2.79 & 9.6$\pm$3.0\phn & 14.096 & 13.204 & 1.925 & 3.633 & 1.708 \\
HSB1140$+$2711 & 2.63 & 8.6$\pm$2.6\phn & 14.372 & 13.169 & 2.201 & 3.517 & 1.316 \\
HSB1141$+$4201 & 2.12 & 8.6$\pm$2.6\phn & 14.274 & 13.211 & 2.103 & 3.605 & 1.502 \\
HSB1310$+$4308 & 2.60 & 10.0$\pm$2.8\phn & 14.131 & 13.237 & 1.960 & 3.667 & 1.707 \\
HSB1337$+$2123 & 2.70 & 6.8$\pm$2.1\phn & 14.213 & 13.062 & 2.042 & 3.419 & 1.377 \\
\noalign{\smallskip}\hline
\noalign{\smallskip}
 & & & & Priddey et al. (2003b) z$>$5 & & & \\
\noalign{\smallskip}\hline
\noalign{\smallskip}
SDSSJ1306$+$0356 & 5.99 & 3.7$\pm$1.0\phn & 13.977 & 12.646 & 1.806 & 2.847 & 1.041 \\
SDSSJ1044$-$0125 & 5.74 & 6.1$\pm$1.2\phn & 14.096 & 12.866 & 1.925 & 3.169 & 1.244 \\
SDSSJ0756$+$4104 & 5.09 & 13.4$\pm$2.1\phn & 13.650 & 13.223 & 1.479 & 3.703 & 2.224 \\
SDSSJ0338$+$0021 & 5.07 & 11.9$\pm$2.0\phn & 13.690 & 13.172 & 1.519 & 3.644 & 2.125 \\
\noalign{\smallskip}\hline
\noalign{\smallskip}
\end{tabular}
\scriptsize
Notes: The columns have the same meanings as those in Table 1 but for 
detected QSOs at $850\mu {\rm m}$. \\ 
$^{a}$The name with prefix RXJ.
\end{table}
\clearpage
\begin{table}[]
\caption[]{Binned data for $850\mu {\rm m}$ non-detected QSOs}
\begin{center}
\begin{tabular}{ccccc}
\hline
\hline\noalign{\smallskip}
Number & mean redshift & weighted mean S$_{\rm 850\mu m}$ & 
weighted mean log$(\frac{L_{\rm bol}}{L_{\odot}})$ & 
weighted mean log$(\frac{L_{60\mu{\rm m}}}{L_{\odot}})$ \\
\noalign{\smallskip}
\hline\noalign{\smallskip}
& & & X-ray absorbed QSOs$^{a}$ & \\
\noalign{\smallskip}
\hline\noalign{\smallskip}
10 & 1.289 & 0.772$\pm$0.354\phn & 12.652 & 12.210$\pm$0.199\phn \\
\noalign{\smallskip}
\hline\noalign{\smallskip}
& & & X-ray unabsorbed QSOs & \\
\noalign{\smallskip}
\hline\noalign{\smallskip}
8 &  1.318 & 0.405$\pm$0.324\phn & 12.854 & 11.930$\pm$0.347\phn \\
9 &  1.840 & 0.717$\pm$0.320\phn & 12.935 & 12.153$\pm$0.194\phn \\
2 &  2.262 & 0.379$\pm$0.669\phn & 12.993 & 11.844$\pm$0.766\phn \\
\noalign{\smallskip}
\hline\noalign{\smallskip}
& & & Isaak et al. (2002) & \\
\noalign{\smallskip}
\hline\noalign{\smallskip}
30 & 4.214 & 2.179$\pm$0.533\phn & 14.224 & 12.468$\pm$0.106\phn \\
\noalign{\smallskip}
\hline\noalign{\smallskip}
& & & McMahon et al. (1999) & \\
\noalign{\smallskip}
\hline\noalign{\smallskip}
3 & 4.463 & 3.167$\pm$0.924\phn & 14.203 & 12.619$\pm$0.127\phn \\
\noalign{\smallskip}
\hline\noalign{\smallskip}
& & & Priddey et al. (2003a) z$>\sim$2 & \\
\noalign{\smallskip}
\hline\noalign{\smallskip}
10 & 1.765 & 0.741$\pm$0.829\phn & 14.176 & 12.173$\pm$0.486\phn \\
27 & 2.190 & 2.547$\pm$0.546\phn & 14.170 & 12.677$\pm$0.093\phn \\
11 & 2.725 & 1.118$\pm$0.871\phn & 14.298 & 12.275$\pm$0.338\phn \\
\noalign{\smallskip}
\hline\noalign{\smallskip}
& & & Priddey et al. (2003b) z$>$5 & \\
\noalign{\smallskip}
\hline\noalign{\smallskip}
6 & 5.100 & 2.078$\pm$0.726\phn & 13.627 & 12.413$\pm$0.152\phn \\
3 & 5.577 & 1.800$\pm$0.884\phn & 13.992 & 12.339$\pm$0.213\phn \\
1 & 6.280 & 1.300$\pm$1.000\phn & 13.979 & 12.189$\pm$0.334\phn \\
\noalign{\smallskip}
\hline\noalign{\smallskip}
\end{tabular}
\end{center}
\scriptsize
Notes: The binned data values for non-detected QSOs (at levels 
of $3\sigma$) at $850\mu {\rm m}$. The columns have the same meanings 
as those in Table 2 except that col(3) is the weighted mean flux 
density at $850\mu {\rm m}$. Note that some redshift bin widths are slightly 
different from 0.5 as the sample redshift range is slightly 
larger or smaller than 0.5. \\ 
$^{a}$One QSO at z=2.46 with a flux density 
 of $-1.6\pm1.2\phn$ at $850\mu {\rm m}$ was not included in the 
estimation because of its large redshift compared to the other non-detected 
QSOs (with $z\leq 1.5$).
\end{table} 
\clearpage
\begin{figure}
\begin{center}
\psfig{figure=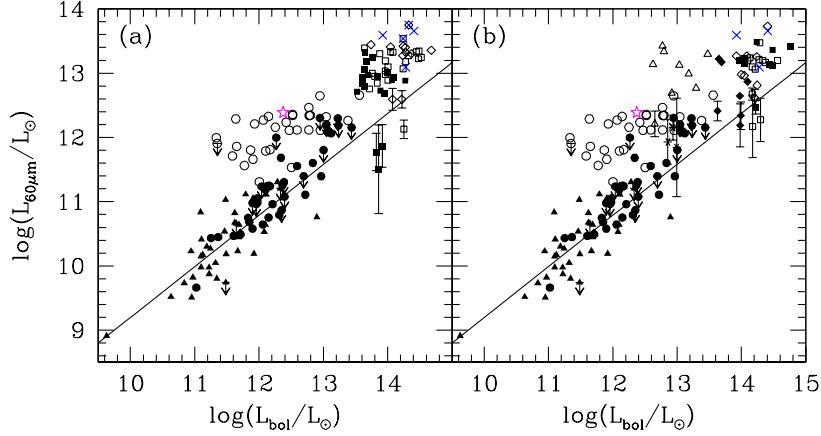,width=120mm}
\caption{The monochromatic luminosity at $60\mu{\rm m}$ vs. the bolometric 
luminosity of AGN for low-z objects and high-z QSOs 
observed (a) in the mm and (b) in the submm. 
In both panel (a) and (b): the open circles represent IR QSOs; the filled 
circles and triangles represent PG QSOs and NLS1s respectively; the crosses 
are for the three QSOs with CO and/or HCN observations; the pentagram 
represents the distant ($z=5.5$) optically faint ($M_{\rm B} \sim -24.2$) 
QSO (see text); the solid line is the best regression line for the local PG 
QSOs and NLS1s obtained from survival analysis; the data points 
with error bars indicate they are the binned values for (sub)mm non-detected 
QSOs (see Sect. 4.1). In panel (a), the open and filled squares 
represent the QSOs 
taken from Omont et al. (2001) and Carilli et al. (2001) respectively; 
the diamonds represent the QSOs from Omont et al. (2003). 
In panel (b), the open triangles 
indicate the X-ray absorbed QSOs taken from Stevens et al. (2005); the 
open and filled squares represent objects from Priddey et al. (2003a) and 
Isaak et al. (2002) respectively; 
the open and filled diamonds indicate the QSOs obtained from McMahon et al. 
(1999) and Priddey et al. (2003b). The stars with error bars are the 
binned data for X-ray unabsorbed, submm non-detected QSOs 
from Page et al. (2004).} 
\end{center}
\end{figure}
\clearpage
\begin{figure}
\begin{center}
\psfig{figure=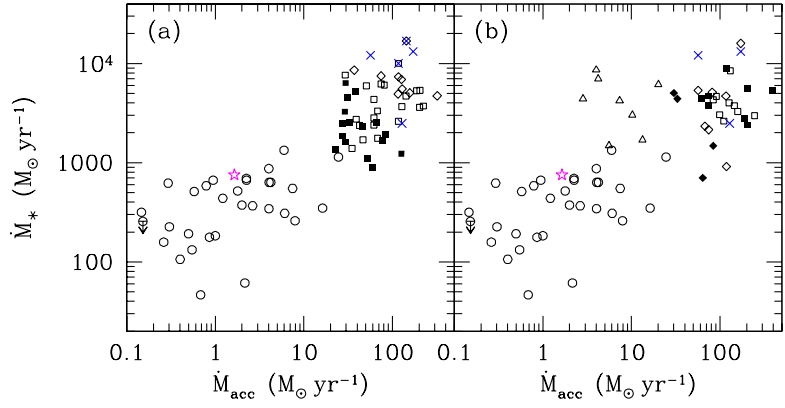,width=120mm}
\caption{The star formation rate ($\Mstardot$) vs. the accretion rate ($\Mdot$) 
for local IR QSOs and 
high-z QSOs detected (a) in the mm and (b) in the sub-mm. 
The symbols are the same as in Fig.~1.}
\end{center}
\end{figure}
\clearpage
\begin{figure}
\begin{center}
\psfig{figure=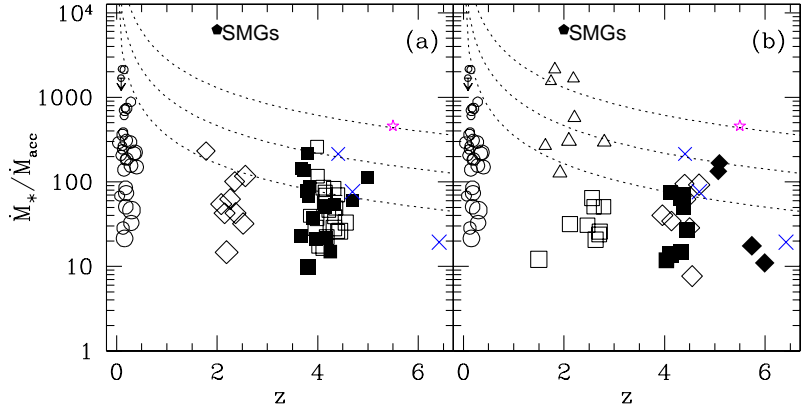,width=120mm}
\caption{
The ratio of $\Mstardot$ and $\Mdot$ vs. redshift for local IR QSOs and (a) 
high-z mm detected QSOs and (b) high-z submm detected QSOs. The 
symbols are the same as in Fig.~1. 
The sizes of the data points indicate their bolometric luminosities 
associated with AGNs. The dotted lines are for three flux limits 
which correspond to a bolometric luminosity of $10^{11}L_\odot$, 
$10^{12}L_\odot$ and $10^{13}L_\odot$ 
(from top to bottom) at redshift 2 respectively. For each curve, 
we assume the relation between the star formation rate and the accretion rate 
by simply fitting a regression line to the data points in Fig.~2a 
 using survival analysis.}
\end{center}
\end{figure}

\end{document}